# Advisors intelligence through sales analytics

A dissertation submitted in partial fulfillment

of the requirements for the degree of

Master of Technology

in

Computer Science & Engineering

with specialization in Big Data

by

Gayatri Pradhan

15MCB1008

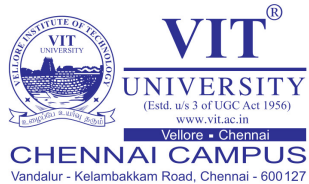

School of Computing Science & Engineering,

VIT University Chennai,

Vandalur-Kelambakkam Road,

Chennai - 600127, India.

May 2017

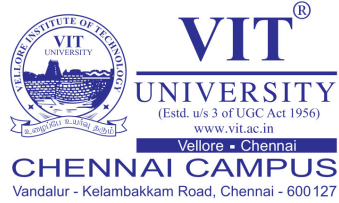

# Declaration

I hereby declare that the dissertation **Advisor Intelligence through Purchase Patterns and Sales Analytics** submitted by me to the School of Computing Science and Engineering, VIT University Chennai, 600 127 in partial fulfillment of the requirements for the award of **Master of Technology** in **Computer Science & Engineering with specialization in Big Data** is a bona-fide record of the work carried out by me under the supervision of **Prof. SyedIbrahim S. P**.

I further declare that the work reported in this dissertation, has not been submitted and will not be submitted, either in part or in full, for the award of any other degree or diploma of this institute or of any other institute or University.

Sign:

Name & Reg. No.:

Date:

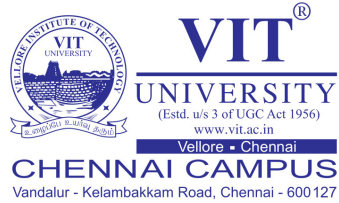

## School of Computing Science & Engineering

# Certificate

This is to certify that the dissertation entitled **Advisor Intelligence through Purchase Patterns and Sales Analytics** submitted by **Gayatri Pradhan** (**Reg. No. 15MCB1008**) to VIT University Chennai, in partial fulfullment of the requirement for the award of the degree of **Master of Technology** in **Computer Science & Engineering with specialization in Big Data** is a bona-fide work carried out under my supervision. The dissertation fulfills the requirements as per the regulations of this University and in my opinion meets the necessary standards for submission. The contents of this dissertation have not been submitted and will not be submitted either in part or in full, for the award of any other degree or diploma and the same is certified.

**Supervisor**  **Program Chair**

Signature:  ....................  Signature:  ....................

Name:  ....................  Name:  ....................

Date:  Date:

**Examiner**

Signature:  ....................

Name:  ....................

Date:

(Seal of the School)

# *Abstract*

In mutual fund, an individual or a firm that is in the business of giving advice about securities to clients is an investment advisor. Investment advisers are individuals or firms that receive compensation for giving advice on investing in stocks, bonds, mutual funds, or exchange traded funds. Investment advisors manage portfolios of securities. Advisors can use new cognitive and analytics capabilities to better understand their clients and their needs and have a stronger ability to deepen relationships with a better portfolio. In this paper, we analyze data points for each advisor, and distinguish the best prospects, obtain insight into their experience and credentials, and learn about their portfolio, in other words to recognize the pattern of portfolio of the advisors. Such analysis helps the sales people to sell the fund company products to the suitable advisors based on the nature of the product they want to sell. This is done by investigating what kind of products advisors have been buying, and what kind of products they might be looking for. This helps to increase the sales of the products as sales people will be reaching the appropriate advisors.

# Acknowledgements

I wish to express my sincere thanks to Dr.G.Viswanathan, Chancellor, Mr. Sankar Viswanathan, Vice President, Ms. Kadhambari S. Viswanathan, Assistant Vice President, Dr. Anand A. Samuel, Vice Chancellor and Dr. P. Gunasekaran, Pro-Vice Chancellor for providing me an excellent academic environment and facilities for pursuing M.Tech. program. I am grateful to Dr. Vaidehi Vijayakumar, Dean of School of Computing Science and Engineering, VIT University, Chennai and to Dr. V. Vijayakumar, Associate Dean. I wish to express my sincere gratitude to Dr.Bharadwaja Kumar G, Program chair of M.Tech CSE with Specialization in Big Data for providing me an opportunity to do my project work. I would like to express my gratitude to my internal guide Prof. SyedIbrahim S. P and my external guide Mr.Inigo Fernando who inspite of their busy schedule guided me in the correct path. I am thankful to Broadridge Financial Solutions, Hyderabad for giving me an opportunity to work on my project and helped me gain knowledge. I thank my family and friends who motivated me during the course of the project work.

# Contents

# List of Figures

# List of Tables

*For/Dedicated to/To my...*

# Chapter 1

# Introduction

The financial services industry was as of late recognized as the business well on the way to be disturbed and changed by millennial in the US. There are comparable signs at the worldwide level. The adjustments in the keeping money and budgetary administrations industry in the coming years will be seismic. For instance, worldwide banks are setting up development focuses and concentrated groups to concentrate on block-chain, proclaimed as a problematic constrain that offers various open doors, for example, redesigning existing keeping money foundation, accelerating settlements and streamlining stock trades.

Financial institutions are in effect consistently tested by contracting incomes and need to enhance operational cost efficiencies. Rising fintech new businesses and officeholder innovation monsters are conveying new plans of action bringing on disturbance and testing customary managing an account plans of action. Controllers in all topographies are requesting stricter consistence and more grounded money related train, from the Federal Reserve in the US to the European Banking Authority (EBA) in the EU, the Prudential Regulation Authority (PRA) and the Financial Conduct Authority (FCA) in the UK.

Financial institutions have the advantages of vast client bases and access to rich value-based information. Making more current business models or frameworks that use the accessible information permits money related foundations to adapt information to convey prevalent client esteem.

Winning in this dynamic market will be supported by how the money related establishments can get an incentive from information. The union of machine and human insight is disturbing customary basic leadership by outfitting associations with learning and knowledge to anticipate and recommend business results. Propels in enormous information and investigation have prompted new items, arrangements and administrations making monetary establishments more brilliant, coordinated and more focused. More up to date administrative and consistence prerequisites, extortion and against tax evasion preventive strides are putting more accentuation on more grounded administration and hazard administration. Information security and information assurance is picking up centrality. This is driving up working costs requiring money related organizations to investigate roads to enhance operational efficiencies. Here are some key patterns reshaping the money related administrations industry:

- To a great degree substantial informational indexes to analyze to reveal patterns, trends and correlations

- Real-time predictive and prescriptive analytics for driving profound significant bits of knowledge

- Hazard and consistence requesting convenient accessibility of dependable, quantifiable and secure information

- Appropriation of machine learning and intellectual abilities

- Democratization of information empowering more self-administration

- Consumerization of BI through best-of-breed information revelation, investigation and visualization tools

- Advanced stages controlled by 360 perspective of clients

- Data and BI scene change and modernization to decrease cost and grasp new-age technologies

- Explanatory ace information administration abilities

- Fortifying information administration capacities

Among financial specialists owning common store offers, the greater part of it holds the reserve shares through a middle person, for example, an agent merchant, bank, subsidize market or stage, insurance agency, speculation consultant. Speculators pick delegate which best suits their necessities.

Financial specialists utilize go-betweens to acquire various advantages. Speculators regularly indicate premiums to the go-between for proposals on the best way to contribute their cash. With the assistance of the venture counsel, the speculator may choose to incorporate common supports as a major aspect of an arrangement of speculations. Despite how resources are distributed between sorts of securities, with the decision to buy common supports, a financial specialist turns into a common reserve shareholder. Because of the vast number of shareholders who want to utilize go-betweens, a mediator crosses over any barrier between common finances and store shareholders.

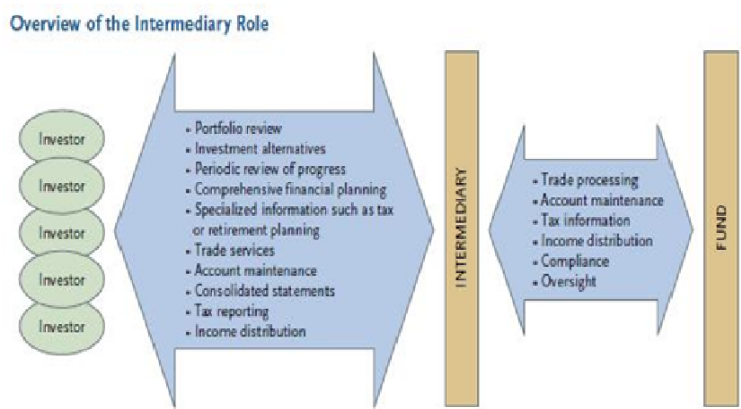

Figure 1.1: Role of Advisors.

Big data analytics helps to learn the portfolios of the investment advisors and recognize the behavioral pattern of investment advisor. Such analysis will help the sales person or the fund companies to find relevant advisor which will be beneficial for the sales person to increase the sales of his products.

Further this recognized pattern can be mapped into ten global broad Morningstar category groups (Equity, Allocation, Convertibles, Alternative, Commodities, Fixed Income, Money Market, Tax Preferred, Property, and Miscellaneous).

The Morningstar Global Category assignments were introduced in 2010 to help investors search for similar investments entertained across the globe. There are different flavors of funds and the investment advisors can be categorized into different categories. Financial time series analysis is worried with hypothesis and routine of benefit valuation after some time. It is a profoundly exact teach, yet like other logical fields hypothesis frames the establishment for making induction. There is, in any case, a key component that recognizes money related time arrangement investigation from other time arrangement examination. Both money related hypothesis and its exact time arrangement contain a component of vulnerability. For instance, there are different meanings of benefit unpredictability, and for a stock return arrangement, the instability is not straightforwardly perceptible. Statistical theory and methods play an important role in financial time series analysis. Our financial time series data is a multivariate time series data set. A Multivariate time series data analysis is used when one wants to model and explain the interaction and co-movements among a group of time series variables. Hence multivariate time series data can be handled by converting it into univariate data using appropriate similarity measures and then building model on it. This can be pictured as shown in the figure Work Flow :

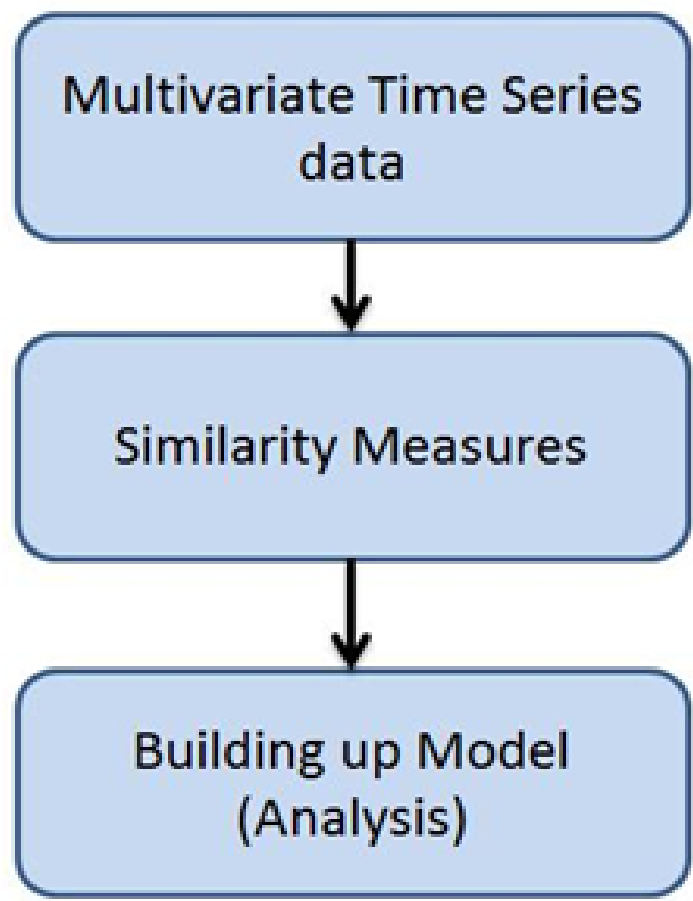

Figure 1.2: Work flow.

This similarity measure can vary from problem to problem. For example, as author [4] proposed a separation work in view of the accepted autonomous Gaussian models and utilized a hierarchical clustering strategy to gathering regularity groupings into an attractive number of clusters. Here an independent Gaussian model is a distance function which is a similarity measure and model is build using hierarchical clustering. Similarly n number of analysis can be done on multivariate time series data. In this paper few of the analysis has been done on multivariate time series data. The first section of the paper explains an approach towards clustering multivariate time series data. This will cluster advisors with similar behavior into same clusters. The second section of the paper explains detecting leaders from correlated advisors among n number of advisors. An approach of analyzing lead-lag relation among a set of time series data is explained.

# Chapter 2

# Literature Survey

In recent past, there is an increased interest in time series clustering research, particularly for finding useful similar trends in multivariate time series in various applied areas such as environmental research, finance, crime, etc. . Clustering multivariate time series has potential for analyzing large volume of finance data at different time points as investors are interested in finding market trends of various funds such as Equity, Allocation, Convertibles, Alternative, Commodities, Fixed Income, Money Market, Tax Preferred, Property, and Miscellaneous so that it will help them to invest.

## 2.1 Multivariate Time Series Clustering Approach for crime Trends Predictions.

In this paper, a novel approach in light of element time wrapping and parametric Minkowski display has been proposed to discover comparable wrongdoing patterns among different wrongdoing groupings of various wrongdoing areas and consequently utilize this data for future wrongdoing patterns forecast. Investigation on Indian wrongdoing records demonstrate that the proposed strategy by and large beats the current strategies in grouping of such multivariate time arrangement information.

## 2.2 Clustering Multivariate Time Series Data

Singhalet. al. [3] has proposed a procedure in view of ascertaining the level of closeness between multivariate timeseries datasets utilizing two similitude elements. One closeness element depends on vital segment investigation and the edges between the main segment subspaces while the other depends on the Mahalanobis remove between the datasets. The philosophy has the essential constraints of PCA and comes up short for the information with no relationship among different measurements.

## 2.3 Time series gene expression data clustering and pattern extraction in Arabidopsis thaliana phosphatise-encoding genes

Creator Pooya Sobhe Bidari [6] exhibited two stage practical grouping as another approach in quality bunching for grouping time arrangement quality expression information. The proposed approach depends on finding utilitarian examples of time arrangement utilizing Fuzzy C-Means and K-implies calculations.

Pearson association equivalence measure is used to isolate the expression cases of characteristics. In this approach, qualities are assembled by K-means and FCM methods according to theirs time course of action expression, then cases of value direct are expelled. By then, new components are described for the qualities and by figuring Pearson association between's new segment vectors, qualities with practically identical expression lead are procured which can incite find interconnections between qualities.

## 2.4 Detecting Climate Change in Multivariate Time Series Data by Novel Clustering and Cluster Tracing Techniques

For identifying environmental change in multivariate information Hardy Kremer [7] proposes novel bunching and grouping following procedures. In this novel grouping approach, time arrangement is part into disjoint, break even with length interims and after that thickness based subsequence bunching methodology is connected, and dynamic time twisting is utilized as a separation measure.

## 2.5 Distance measures for effective clustering of ARIMA time-series

K. Kalpakis [37] concentrated the grouping of ARIMA time course of action, by using the Euclidean partition between the Linear Predictive Coding cepstra of two time-plan as their uniqueness measure. The cepstral coefficients for an AR(p) time course of action are gotten from the auto-backslide coefficients. The package around medoids methodology that is a k-medoids computation was picked, with the gathering comes to fruition evaluated with the cluster closeness measure and Silhouette width. In light of a trial of four educational files, they showed that the LPC cestrum gives higher unjustifiable vitality to uncover to one time course of action from another and predominant gathering than other comprehensively used methods, for instance, the Euclidean detachment between (the underlying 10 coefficients of) the DFT, DWT, PCA, and DFT of the auto-relationship limit of two time game plan.

## 2.6 Stock markets forecasting based on fuzzy time series model

Yupei Lin [41] attempted to enhance the forecast precision with amending two inadequacies, sub interims neglecting to well speak to the information appropriation structures and a solitary precursor figure the fluffy relationship in current fluffy time arrangement display. To begin with, he distributed universe of talk in subintervals with the midpoints of two neighboring gatherings centers, and the subintervals are used to fuzzily the time course of action into cushy time plan. At that point, the fluffy time arrangement display with multi calculates high request fluffy relationship is developed to conjecture the share trading system. The outcomes demonstrated that the model enhanced the expectation exactness when contrasted and the benchmark show.

## 2.7 Mixtures of ARMA Models for Model-Based Time Series Clustering

Xiong and Yeung [38] proposed a model-based method for gathering univariate ARIMA course of action. They expected that the time course of action are created by k unmistakable ARMA models, with each model identifies with one gathering of interest. A desire expansion (EM) calculation was utilized to take in the blending coefficients and additionally the parameters of the part models that boost the desire of the total information log-probability. What's more, the EM calculation was enhanced so that the quantity of groups could be resolved consequently.

## 2.8 Risk analysis of China stock market based on VAR model

This paper is essentially from the meaning of VAR model, as VAR model is one the most imperative approaches to gauge advertise chance. In light of VAR model Xinjie Ma and Yongsheng Yang [41] proposed a way to deal with anticipate the danger of China's securities exchange by means of observational examination made on Shanghai Index and picks five examples for portfolio hazard look into so it has incredible centrality to China's stock exchange chance measure. The investigation of the venture chances in the share trading system gives an incredible reference an incentive to the financial specialists.

# Chapter 3

# Experimental Design & Setup

## 3.1 Experimental Design

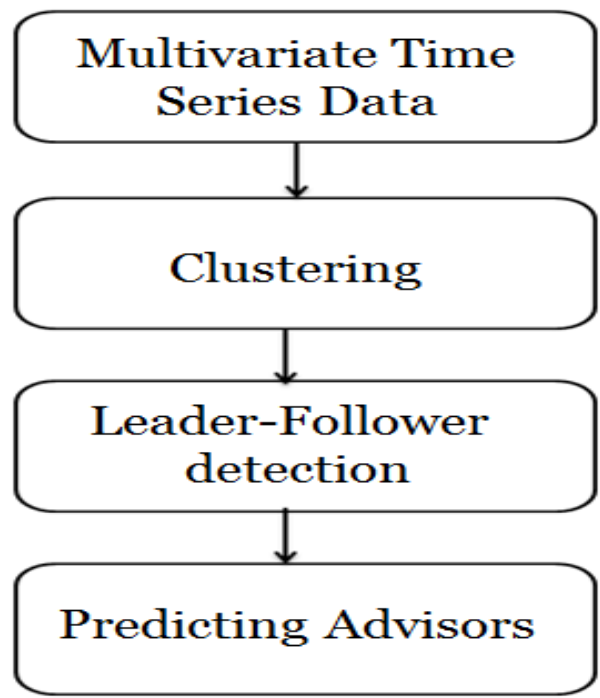

Figure 3.1: Methodology of predicting Advisors behavior

## 3.2 Module 1: Clustering similar advisors

A novel approach based on cosine similarity with hierarchical clustering model has been proposed to find similar advisors.

The Morningstar Category groupings were acquainted in 1996 with help speculators make important correlations between mutual funds. Morningstar found that the speculation objective recorded in a reserve's plan regularly did not satisfactorily clarify how the store really contributed. For instance, many assets asserted to look for "development," however some of those were putting resources into set up blue-chip organizations while others were putting resources into little top organizations. The Morningstar Category characterizations tackled this issue by breaking portfolios into associate gatherings in view of their holdings. The classifications help investors distinguish the top-performing funds, evaluate potential risk, and construct very much well-diversified portfolios. Morningstar routinely surveys the classification structure and the portfolios inside every classification to guarantee that the framework addresses the issues of speculators. Morningstar doles out classes to a wide range of portfolios, for example, common assets, variable annuities, and separate records. Portfolios are set in a given classification in view of their normal possessions insights in the course of recent years. Morningstar's article group additionally audits and supports all class assignments. On the off chance that the portfolio is new and has no history, Morningstar gauges where it will fall before giving it a more lasting classification task. Whenever vital, Morningstar may change a class task in view of late changes to the portfolio.

In clustering, Euclidean distance measure is the most commonly used for non-time series data clustering. While working with financial multivariate time series data, it is not suitable for multivariate time series clustering. Instead of Euclidean distance measure in stand-alone mode, cosine similarity provides better results. The problem of finding similar advisor can be solved in two steps:

1. Hierarchical clustering.
2. Computing cosine similarity

We now discuss each step in detail.

### 3.2.1 Hierarchical Clustering

Clustering can be viewed as the most vital unsupervised learning issue; along these lines, as each other issue of this kind, it manages finding a structure in an accumulation of unlabeled data. A free meaning of clustering could be "the way toward sorting out items into gatherings whose individuals are comparative somehow". A clustering is in this manner a gathering of articles which are "comparable" amongst them and are "dissimilar" to the items having a place with different groups.

A hierarchical clustering algorithm creates a hierarchical decomposition of the given data set objects. Depending on the decomposition approach, hierarchical algorithms are classified as agglomerative (merging) or divisive (splitting). The agglomerative approach starts with each data point in a separate cluster or with a certain large number of clusters. Each step of this approach merges the two clusters that are the most similar. Thus after each step, the total number of clusters decreases. This is repeated until the desired number of clusters is obtained or only one cluster remains. By contrast, the divisive approach starts with all data objects in the same cluster. In each step, one cluster is split into smaller clusters, until a termination condition holds. Agglomerative algorithms are more widely used in practice. In this paper we are using agglomerative algorithm. Agglomerative approach can be done in two different ways: top-down or bottom-up approach. This will result into cluster with advisors of similar type in same cluster. Hence it will result into discovery of similar type of advisors.

Thus after each step, the total number of clusters decreases. This is repeated until the desired number of clusters is obtained or only one cluster remains. By contrast, the divisive approach starts with all data objects in the same cluster. In each step, one cluster is split into smaller clusters, until a termination condition holds. Agglomerative algorithms are more widely used in practice. In this paper we are using agglomerative algorithm. Agglomerative approach can be done in two different ways: top-down or bottom-up approach. This will result into cluster with advisors of similar type in same cluster. Hence it will result into discovery of similar type of advisors.

### 3.2.2 Cosine similarity

Using cosine similarity, advisor with strong similarity can be evaluated. Later, this will help to detect the leader and the follower among the strongly similar advisors.

Since similar behavior advisors are bucketed into one cluster, a sample portfolio from each cluster is considered to compute the cosine similarity among each advisors.

Cosine similarity is a measure of comparability between two non-zero vectors of an internal item space that measures the cosine of the edge between them. The cosine of 0 is 1, and it is under 1 for some other edge. It is in this manner a judgment of introduction and not size: two vectors with a similar introduction have a cosine closeness of 1, two vectors at 90 have a comparability of 0, and two vectors oppositely contradicted have a similitude of - 1, autonomous of their greatness. Cosine closeness is especially utilized as a part of positive space, where the result is perfectly limited in [0,1]. The name gets from the expression "heading cosine": for this situation, take note of that unit vectors are maximally "comparative" on the off chance that they're parallel and maximally "different" on the off chance that they're orthogonal (opposite). This is practically equivalent to the cosine, which is solidarity (maximum value) when the portions subtend a zero edge and zero (uncorrelated) when the fragments are opposite. The technique is additionally used to quantify cohesion within clusters.

Consider an example to find the cosine similarity between two time series t1 and t2 which describes the portfolio of the advisors.

t1 =(5, 0, 3, 0, 2, 0, 0, 2, 0, 0)

t2 =(3, 0, 2, 0, 1, 1, 0, 1, 0, 1)

Calculate cosine similarity using the following formula:

$$cos(t1, t2) = 1.2/(|1||2|) = 0.94$$

Similarly find cosine similarity between each pair of time series data. This cosine score is then stored in a database table.

Table 3.1: Cosine Similarity table

|    | **t1**    | **t2**    | **t3**    | **t4**    |
|----|-----------|-----------|-----------|-----------|
| **t1** | cos(t1,t1) | cos(t1,t2) | cos(t1,t3) | cos(t1,t4) |
| **t2** | cos(t2,t1) | cos(t2,t2) | cos(t2,t3) | cos(t2,t4) |
| **t3** | cos(t3,t1) | cos(t3,t2) | cos(t3,t3) | cos(t3,t4) |
| **t4** | cos(t4,t1) | cos(t4,t2) | cos(t4,t3) | cos(t4,t4) |

This result will give group or pairs of advisors whose portfolio behaviors are highly correlated.

## 3.3 Module 2: Leader-Follower detection

This outcome will give gathering or matches of counsels whose portfolio practices are profoundly associated.

An augmentation to above got comes about prompted find pioneers among various time arrangement information. This issue of recognizing pioneer among guide with comparable portfolio is comprehended by breaking down lead-slack relations among the time arrangement information. A proficient calculation has been clarified by creators [2] which can track the slacked relationship and process the pioneers incrementally is been proposed for atmosphere science information. The issue of authority revelation can be settled in three fundamental strides:

1. Compute the slacked relationship (cross-connection) between each combine of time arrangement;

2. Construct an edge-weighted guided chart in light of slacked connections to dissect the lead-slack connection among the arrangement of time arrangement.

3. Detect the pioneers by examining the initiative transmission in the chart. We now discuss each step in detail.

### 3.3.1 Lagged Correlation Computation

The initial phase in identifying pioneer among set to time arrangement information is to process the slacked connection between's each match of time arrangement.

We propose to add up to the effects of various slacks and describe an amassed slacked association. The amassed slacked relationship figuring can be lit up by the running with case. Fig. 2(a) exhibits two time game-plan X (top) and Y (base) with a length of 150 that recommends time learn of 150 x-focus and the time approach variable at y-focus. The window length is set to be 120 and the window set apart by the specked rectangle. Fig. 2(b) shows the slacked relationship among's X and Y at each slack l enrolled by Eq. (1) over the two windows. The condition is according to in figure 3.2:

$$\rho_{t,w}^{ij}(l) = \begin{cases} \frac{\sum_{\tau=t-w+1}^{t-l}(s_{\tau+l}^i - \overline{s_{t,w-l}^i})(s_\tau^j - \overline{s_{t-l,w-l}^j})}{\sigma_{t,w-l}^i \sigma_{t-l,w-l}^j}, & l \geq 0; \\ \rho_{t,w}^{ji}(-l), & l < 0, \end{cases}$$

FIGURE 3.2: The lagged correlation Equation

The most extreme slack m = 60, i.e., mod(l) 60. At the point when the processed slack l ¡ 0 (that is considered as Y is deferred by X with slack l), the positive relationship exists for l [60, 39] (the shadowed range). At the point when l 0 (i.e., X is deferred by Y with slack l), beginning from l = 1, we can watch a solid increment in positive connection and it accomplishes a pinnacle estimation of 0.81 at l = 32. We have to total all the watched relationship values over the whole slack traverse keeping in mind the end goal to recognize the administration (X drives Y or Y drives X), and take the normal connection esteem given the two instances of l. The amassed slacked connection between's two time arrangement Si and Sj, signified as Eij, is then characterized as the bigger expected relationship esteem:

$$Eij = \max\left(Eij\,(r|l{\geq}0),\,Eij\,(r|l{<}0)\right)$$

FIGURE 3.3: Aggregated lagged correlation between two time series

We say that Si leads Sj if l is less than 0, and Si led by Sj otherwise if l is greater than 0. Such leadership (Si leads Sj or vice versa depending on l value) is also called the lead-lag relation between Si and Sj.

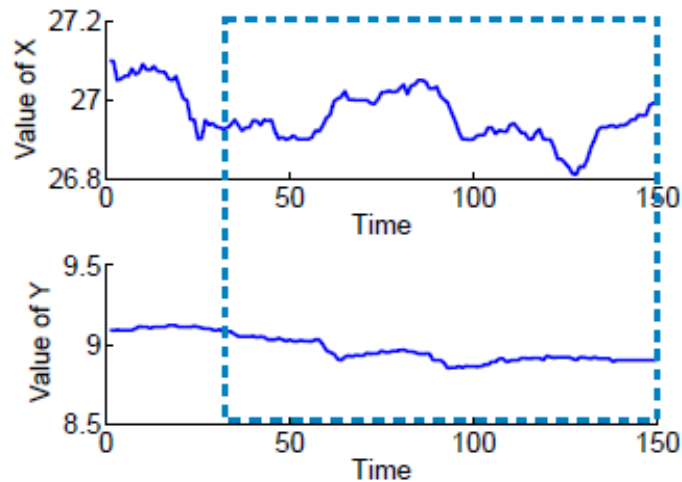

FIGURE 3.4: Two Time Series

In light of this figured lead-slack connection esteem we can build a chart. This will separate pioneer among set of time arrangement. Subsequently next stride is to build edge-weighted guided chart in view of slacked connections to dissect the lead-slack connection among the arrangement of time arrangement.

### 3.3.2 Graph Construction

With a specific end goal to outline the administration connections among an arrangement of time arrangement, building an edge-weighted diagram will learn lead-slack relationship among the arrangement of time arrangement information.

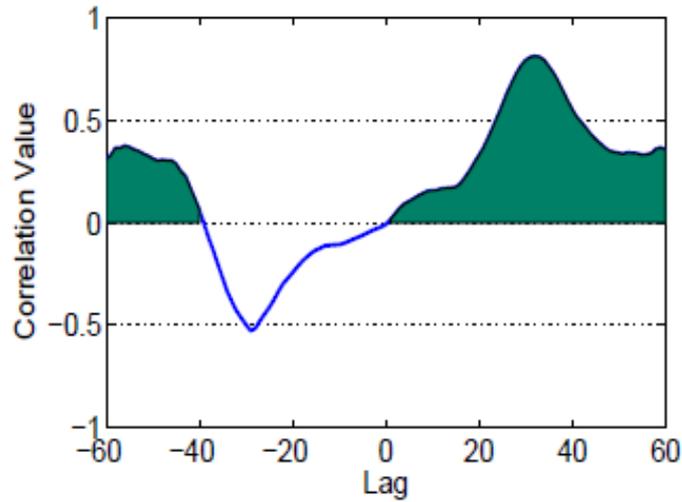

Figure 3.5: Lagged Correlation

Straightforward edge-weighted coordinated chart, G (V, E), where the hubs V = S1, S2 . SN speaks to N time arrangement, and the coordinated edges E speaks to lead-slack relations between combine of time arrangement. An edge (Si, Sj) shows that Si is driven by Sj and its weight is set as Ei j(r). Since we are keen on pivotal lead-slack relations, a connection edge g is set. The edge will reinforce the development of a chart to such an extent that exclusive those sets Si and Sj with Ei j(r) ¿ g have edges in G. This will help in extricating solid pioneer among the time arrangement information by setting the suitable limit esteem.

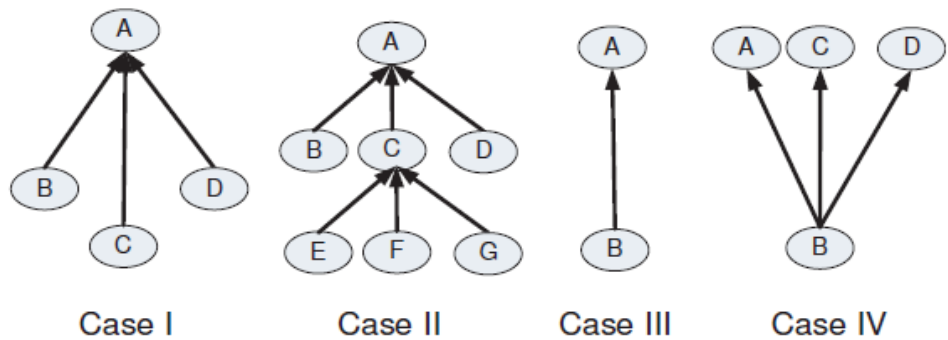

Figure 3.6: Comparison of Leadership Score on Different Graph Structures

### 3.3.3 Leader Extraction

In perspective of the structure of G and the PageRank estimations of time course of action, remove the pioneers by discarding unsuitable specialists. The primitive believed is to first sort the time plan by the sliding solicitation of their PageRank values and after that to oust iteratively the time course of action that is driven either by officially found pioneers or by the relative of previously found pioneers. This will realize number of pioneers.

## 3.4 Module 3: Predicting Advisors Performance

For many years theorists have attempted to make a fiscal benefit in monetary showcases by foreseeing the future cost of wares, stocks, remote trade rates and all the more as of late prospects and alternatives. In the course of the most recent couple of decades these endeavors have expanded particularly, utilizing an assortment of procedures (Hsu, which can be comprehensively ordered into three classes:

- fundamental analysis
- technical analysis
- traditional time series forecasting

### 3.4.1 Fundamental Analysis

Fundamental analysis makes utilization of essential market data keeping in mind the end goal to anticipate future developments of a benefit. On the off chance that a financial specialist was taking a gander at a specific stock's basic information they would consider data, for example, income, benefit estimates, supply, request and working edges and so on. Theorists taking a gander at commodities should seriously mull over weather patterns, political angles, government enactment and etc. Viably fundamental analysis is worried with full scale monetary and political components that may influence the future cost of a money related resource. Basic investigation is not viewed as further in this review.

### 3.4.2 Technical Analysis

Technical analysis is the investigation of verifiable costs and examples with the point of anticipating future costs. Experts of specialized investigation in the past were alluded to as chartists, as they trusted every one of that was had to think about a specific market was contained in its evaluating diagram.Technical analysis (TA) is intriguing as it has a tendency to captivate assessment as to its logical premise and viability.

However, in the realm of finance technical analysis is universal and broadly utilized. In support of TA a plenty of purported pointers have been produced throughout the years from basic moving midpoints to a great deal more colorful offerings. Today every bit of programming or on-line investigation device gives the capacity to put a huge number of specialized markers on a diagram of a stock, item or any money related instrument.

### 3.4.3 Time Series Forecasting

Financial time series data are a sequence of prices of some financial assets over a specific period of time. Fundamental analysis is the examination of the underlying forces that affect the well-being of the economy, industry sectors, and individual companies. For example, in our problem to forecast the future performance of the individual advisors. Here we are forecasting the performance of multiple leading advisors of the financial market. Hence, the data used is multivariate time series data. Since its is a multivariate time series data this problem can be solved using VAR models (vector autoregressive models).

VAR model is used for multivariate time series. The vector autoregression (VAR) model is a standout amongst the best, adaptable, and simple to utilize models for the investigation of multivariate time arrangement. It is a characteristic augmentation of the univariate autoregressive model to dynamic multivariate time arrangement. The VAR display has turned out to be particularly valuable for portraying the dynamic conduct of monetary and money related time arrangement and for estimating. It frequently gives better conjectures than those from univariate time arrangement models and expounds hypothesis based synchronous conditions models. Conjectures from VAR models are very adaptable on the grounds that they can be made restrictive on the potential future ways of indicated factors in the model.

Before proceeding with VAR model, modeling a time series requires certain criteria to satisfy and it includes stationary series, random walks , rho Coefficient, Dickey Fuller Test of Stationarity. I took up this segment first because that until unless your time arrangement is stationary, you cannot build a time series model. In situations where the stationary measure are damaged, the principal imperative moves toward becoming to stationarize the time arrangement and afterward attempt stochastic models to forecast this time arrangement. There are different methods for bringing this stationarity. Some of them are De-trending, Differencing and so on. There are three basic criterion for a series to be classified as stationary series :

1. Constant mean across time (Figure 3.7)

2. Constant variance across time (Figure 3.8)

3. Constant auto-covariance across time (Figure 3.9)

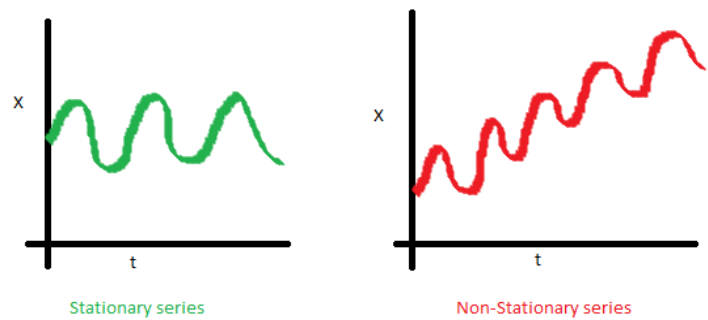

FIGURE 3.7: Constant mean across time

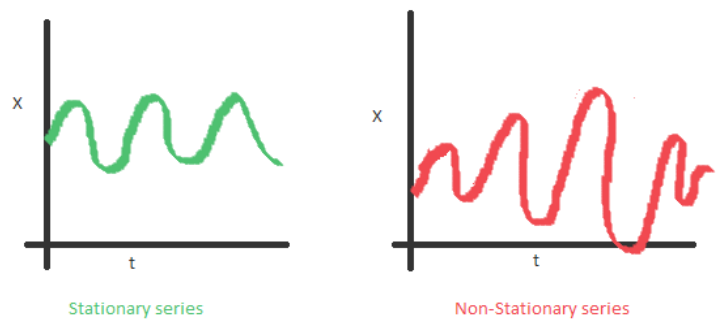

FIGURE 3.8: Constant variance across time

In statistics and econometrics, an Augmented DickeyFuller test (ADF) tests the null hypothesis of a unit root is available in a period arrangement test. The option hypothesis is distinctively relying upon which rendition of the test is utilized, yet is normally stationarity or pattern stationarity. It is an increased adaptation of the DickeyFuller test for a bigger and more entangled arrangement of time arrangement models. The augmented DickeyFuller (ADF) measurement, utilized as a part of the test, is a negative number. The more negative it is, the more grounded the dismissal of the hypothesis that there is a unit root at some level of certainty. Once the series is found stationary, the model can be build on the multivariate time series data.

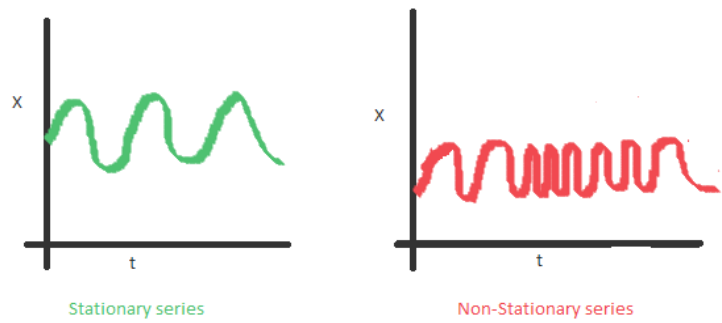

FIGURE 3.9: Constant auto-covariance across time

The structure of VAR is that every variable is a straight capacity of past slacks of itself and past slacks of alternate factors.

As an example suppose that we measure three different time series variables, denoted by

$$x_(t,1), x_(t,2), and x_(t,3)$$

The vector autoregressive model of order 1, denoted as VAR(1), is as follows:

$$x_{t,1} = \alpha_1 + \emptyset_{1,1} x_{t-1,1} + \emptyset_{1,2} x_{t-1,2} + \emptyset_{1,3} x_{t-1,3} + \omega_{t,1}$$

$$x_{t,2} = \alpha_2 + \emptyset_{2,1} x_{t-1,1} + \emptyset_{2,2} x_{t-1,2} + \emptyset_{2,3} x_{t-1,3} + \omega_{t,2}$$

$$x_{t,3} = \alpha_1 + \emptyset_{3,1} x_{t-1,1} + \emptyset_{3,2} x_{t-1,2} + \emptyset_{3,3} x_{t-1,3} + \omega_{t,3}$$

FIGURE 3.10: Vector autoregressive model of order 1

Each variable is a linear function of the lag 1 values for all variables in the set.

In a VAR(2) model, the slack 2 values for all factors are added to the correct sides of the conditions, For the situation of three x-factors (or time series) there would be six indicators on the correct side of every condition, three slack 1 terms and three slack 2 terms.

In general, for a VAR(p) model, the principal p slacks of every variable in the framework would be utilized as regression indicators for every variable. The value of p can be calculated using VARselect method. This helps to select lag value between two time series and further proceed with the VAR model as discussed earlier.

Generally, there are two stages to make and to create a precise forecasting. The initial step is to gather the information that is important to the intended purpose of forecasting and related to information. The second step is to pick the correct forecasting technique to be actualized.

The VAR analysis procedure are shown as follows:

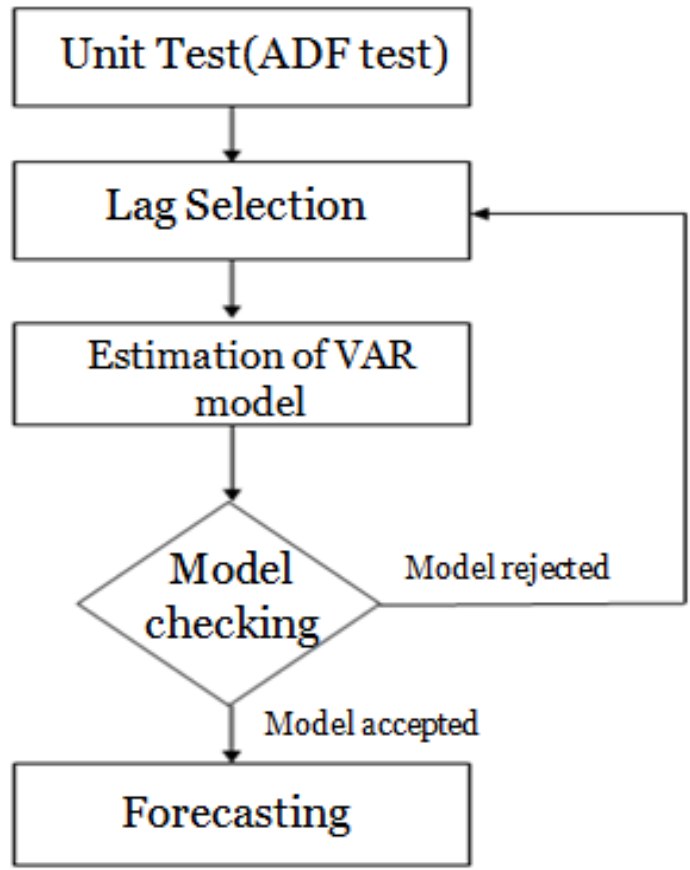

FIGURE 3.11: VAR Model

The underlying stride in the investigation is an examination stasioneritas data. To look at the stationary of the information that can be utilized as a unit root test. The unit root test utilized depends on the Augmented Dickey-Fuller () test. Numerically, the type of ADF is shown as follows:

$$\Delta Y_t = \gamma + \delta t + \rho Y_{t-1} + \sum_{j=1}^{k} \phi_j \Delta Y_{t-j} + e_t$$

where $\Delta Y_t = Y_t - Y_{t-1}$ and $\rho = a - 1$.

FIGURE 3.12: ADF test

The hypothesis of adf test is $H_0$: $\rho = 0$ (a unit root exists).

At the hugeness level of (1 - $\alpha$)100%, $H_0$ is rejected, if the measurements is not as much as the basic incentive at the time of $\alpha$, or p value and not as much as the criticalness estimation of $\alpha$. Implying that, the information is stationary.

The following stride is the determination of the slack request. The points of this progression is to get the ideal slack request of the model. Slack request choice uses the accompanying information criteria(Figure 3.13).

Akaike Information Criterion (*AIC*)

$$AIC(p) = \log \det \left( \hat{\Sigma}_u(p) \right) + \frac{2}{T} pk^2$$

Schwarz Information Criterion (*SC*)

$$SC(p) = \log \det \left( \hat{\Sigma}_u(p) \right) + \frac{\log(T)}{T} pk^2$$

Hannan-Quinn Criterion (*HQ*)

$$HQ(p) = \log \det \left( \hat{\Sigma}_u(p) \right) + \frac{2 \log \log(T)}{T} pk^2$$

FIGURE 3.13: Information criteria for lag order selection

Where, $p$ is lag, $k$ is the number of endogenous variable. Estimation of slack p picked as the estimation of $p^*$ which limits the data criteria in the watched interims of 1, ..., $p_{max}$. Slack is ideal in view of the littlest estimation of *AIC*, *SC* and *HQ*.

The successive altered likelihood ratio (LR) test is completed. Beginning with the greatest slack, trailed by trial of the hypothesis to check whether the coefficients on slack p are mutually zero utilizing the $X^2$ measurements.

After the estimation of the model in view of the ideal slack order, indicative checking for the lingering was done. It means to whether there is a serial connection (autocorrelation) in slack $h$ on residuals

*VAR* model is alluded to the best model, if the model meets the VAR analysis methods that is portrayed previously. The accompanying stride is forecasting of the future periods utilizing the best of *VAR* model. By and large, Mean Square Error (*MSE*) esteem is utilized to decide the precision of estimating results. The type of *MSE* is appeared as:

$$\text{MSE} = \frac{1}{n} \sum_{i=1}^{n} (Y_t - y_t)^2$$

Chapter 3. *Experimental Design & Setup* 23where $n$ is measure of information. A decent model will deliver smallest of esteem which is identified with the precision of a forecast. In a few cases, the qualities can be utilized to decide the execution of a model.

# Chapter 4

# Experiments & Results

## 4.1 Similar Advisors Clustering

The Morningstar Category classifications breaks portfolios into peer groups based on their holdings.Advisors can be clustered based on the their holdings.Portfolios are placed in a category based on their average holdings statistics over the past three years. Here advisors with similar category are clustered into a cluster.

Below is the sample cluster of advisors with similar portfolio.Advisor at the center represents as the leading advisor among the remaining advisors in the cluster.

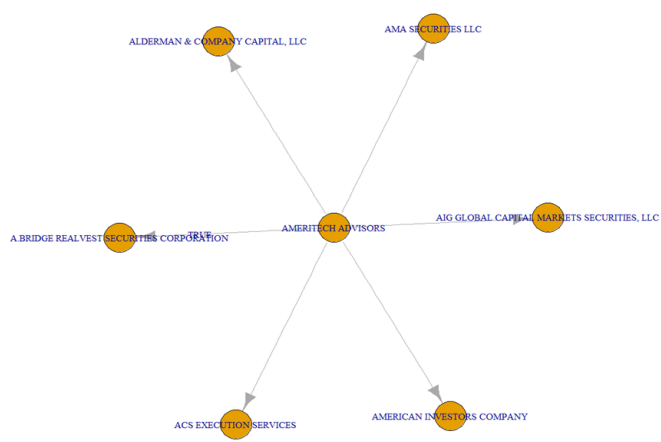

Figure 4.1: Sample Cluster of Advisor with Similar portfolio

## 4.2 Predicting Leading Advisors

Leading advisors from each cluster are extracted from above step.

Below figures are forecast results of sample leading advisors.

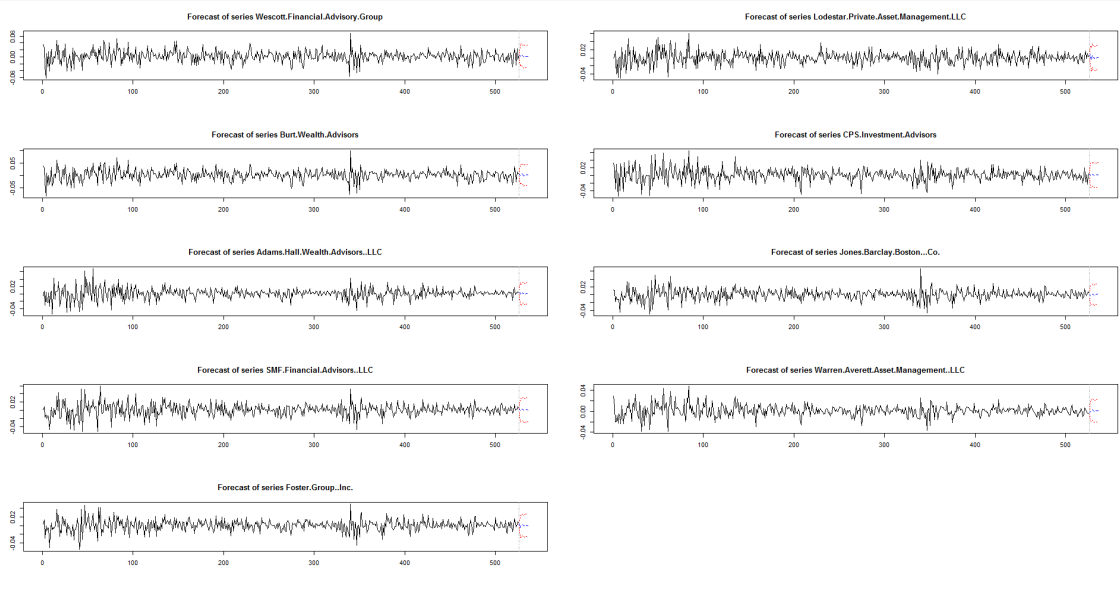

FIGURE 4.2: Prediction using VAR model

```
> predict
$Wescott.Financial.Advisory.Group
              fcst        lower      upper         CI
 [1,]  0.0064698634 -0.02435374 0.03729347 0.03082361
 [2,]  0.0030144518 -0.02923439 0.03526330 0.03224885
 [3,]  0.0007562582 -0.03168867 0.03320119 0.03244493
 [4,]  0.0012204118 -0.03128133 0.03372215 0.03250174
 [5,]  0.0019310376 -0.03058083 0.03444291 0.03251187
 [6,]  0.0015294238 -0.03098494 0.03404378 0.03251436
 [7,]  0.0015127425 -0.03100311 0.03402859 0.03251585
 [8,]  0.0016182383 -0.03089771 0.03413419 0.03251595
 [9,]  0.0015813771 -0.03093469 0.03409745 0.03251607
[10,]  0.0015631004 -0.03095299 0.03407919 0.03251609

$Burt.Wealth.Advisors
              fcst        lower      upper         CI
 [1,]  0.0073280104 -0.03084185 0.04549787 0.03816986
 [2,]  0.0022935079 -0.03943197 0.04401898 0.04172547
 [3,]  0.0003078127 -0.04180508 0.04242071 0.04211290
 [4,]  0.0011157906 -0.04110760 0.04333918 0.04222339
 [5,]  0.0019882858 -0.04025354 0.04423011 0.04224183
 [6,]  0.0014102699 -0.04083492 0.04365546 0.04224519
 [7,]  0.0014287033 -0.04081910 0.04367651 0.04224780
 [8,]  0.0015675490 -0.04068045 0.04381555 0.04224800
 [9,]  0.0015064660 -0.04074171 0.04375464 0.04224818
[10,]  0.0014851138 -0.04076310 0.04373333 0.04224821

$Adams.Hall.Wealth.Advisors..LLC
              fcst        lower      upper         CI
 [1,]  0.0017550232 -0.02597861 0.02948865 0.02773363
 [2,] -0.0013900008 -0.02949243 0.02671243 0.02810243
 [3,]  0.0002582391 -0.02803781 0.02855429 0.02829605
 [4,]  0.0009829520 -0.02735259 0.02931849 0.02833554
 [5,]  0.0005172775 -0.02782181 0.02885636 0.02833908
 [6,]  0.0006096794 -0.02773444 0.02895380 0.02834412
 [7,]  0.0007337002 -0.02761086 0.02907826 0.02834456
 [8,]  0.0006555224 -0.02768931 0.02900036 0.02834484
 [9,]  0.0006401494 -0.02770477 0.02898506 0.02834491
[10,]  0.0006678136 -0.02767711 0.02901274 0.02834492

$SMF.Financial.Advisors..LLC
              fcst        lower      upper         CI
 [1,]  0.0017775763 -0.02656151 0.03011666 0.02833908
 [2,]  0.0004973907 -0.02866522 0.02966001 0.02916261
 [3,] -0.0001640587 -0.02941842 0.02909030 0.02925436
 [4,]  0.0004646594 -0.02884994 0.02977925 0.02931459
 [5,]  0.0010008455 -0.02832074 0.03032243 0.02932159
 [6,]  0.0007052461 -0.02861861 0.03002910 0.02932385
 [7,]  0.0006552201 -0.02866969 0.02998013 0.02932491
 [8,]  0.0007516798 -0.02857332 0.03007668 0.02932500
 [9,]  0.0007271882 -0.02859791 0.03005229 0.02932510
[10,]  0.0007083846 -0.02861673 0.03003350 0.02932511
```

FIGURE 4.3: Prediction using VAR model

```
VAR Estimation Results:
========================
Endogenous variables: Wescott.Financial.Advisory.Group, Burt.Wealth.Advisors, Adams.Hall.Wealth.Advisors..LLC, SMF.Financial.A
dvisors..LLC, Foster.Group..Inc., Lodestar.Private.Asset.Management.LLC, CPS.Investment.Advisors, Jones.Barclay.Boston...Co.,
Warren.Averett.Asset.Management..LLC
Deterministic variables: const
Sample size: 524
Log Likelihood: 16419.35
Roots of the characteristic polynomial:
0.5353 0.5353 0.4199 0.3618 0.3618 0.3466 0.3466 0.3208 0.3208 0.302 0.302 0.3007 0.3007 0.2657 0.2351 0.2351 0.1118 0.1118
Call:
VAR(y = i, p = 2)

Estimation results for equation Wescott.Financial.Advisory.Group:
=================================================================
Wescott.Financial.Advisory.Group = Wescott.Financial.Advisory.Group.l1 + Burt.Wealth.Advisors.l1 + Adams.Hall.Wealth.Advisors.
.LLC.l1 + SMF.Financial.Advisors..LLC.l1 + Foster.Group..Inc..l1 + Lodestar.Private.Asset.Management.LLC.l1 + CPS.Investment.A
dvisors.l1 + Jones.Barclay.Boston...Co..l1 + Warren.Averett.Asset.Management..LLC.l1 + Wescott.Financial.Advisory.Group.l2 + B
urt.Wealth.Advisors.l2 + Adams.Hall.Wealth.Advisors..LLC.l2 + SMF.Financial.Advisors..LLC.l2 + Foster.Group..Inc..l2 + Lodesta
r.Private.Asset.Management.LLC.l2 + CPS.Investment.Advisors.l2 + Jones.Barclay.Boston...Co..l2 + Warren.Averett.Asset.Manageme
nt..LLC.l2 + const

                                          Estimate Std. Error t value Pr(>|t|)
Wescott.Financial.Advisory.Group.l1      -0.049575   0.148336  -0.334   0.7384
Burt.Wealth.Advisors.l1                   0.093539   0.131954   0.709   0.4787
Adams.Hall.Wealth.Advisors..LLC.l1        0.210776   0.084851   2.484   0.0133 *
SMF.Financial.Advisors..LLC.l1            0.109904   0.141152   0.779   0.4366
Foster.Group..Inc..l1                     0.078722   0.181749   0.433   0.6651
Lodestar.Private.Asset.Management.LLC.l1 -0.057414   0.069666  -0.824   0.4103
CPS.Investment.Advisors.l1                0.328977   0.080573   4.083 5.17e-05 ***
Jones.Barclay.Boston...Co..l1            -0.525055   0.258462  -2.031   0.0427 *
Warren.Averett.Asset.Management..LLC.l1  -0.248137   0.151433  -1.639   0.1019
Wescott.Financial.Advisory.Group.l2      -0.009287   0.142938  -0.065   0.9482
Burt.Wealth.Advisors.l2                   0.057965   0.123496   0.469   0.6390
Adams.Hall.Wealth.Advisors..LLC.l2        0.103697   0.089581   1.158   0.2476
SMF.Financial.Advisors..LLC.l2           -0.010637   0.142297  -0.075   0.9404
Foster.Group..Inc..l2                     0.174253   0.179351   0.972   0.3317
Lodestar.Private.Asset.Management.LLC.l2  0.023862   0.062556   0.381   0.7030
CPS.Investment.Advisors.l2                0.018825   0.082690   0.228   0.8200
Jones.Barclay.Boston...Co..l2            -0.186018   0.256896  -0.724   0.4693
Warren.Averett.Asset.Management..LLC.l2  -0.161263   0.147456  -1.094   0.2746
const                                     0.001454   0.000710   2.048   0.0411 *
---
Signif. codes:  0 '***' 0.001 '**' 0.01 '*' 0.05 '.' 0.1 ' ' 1

Residual standard error: 0.01573 on 505 degrees of freedom
Multiple R-Squared: 0.1013,  Adjusted R-squared: 0.06922
F-statistic: 3.161 on 18 and 505 DF,  p-value: 1.399e-05
```

Figure 4.4: VAR Estimation Results

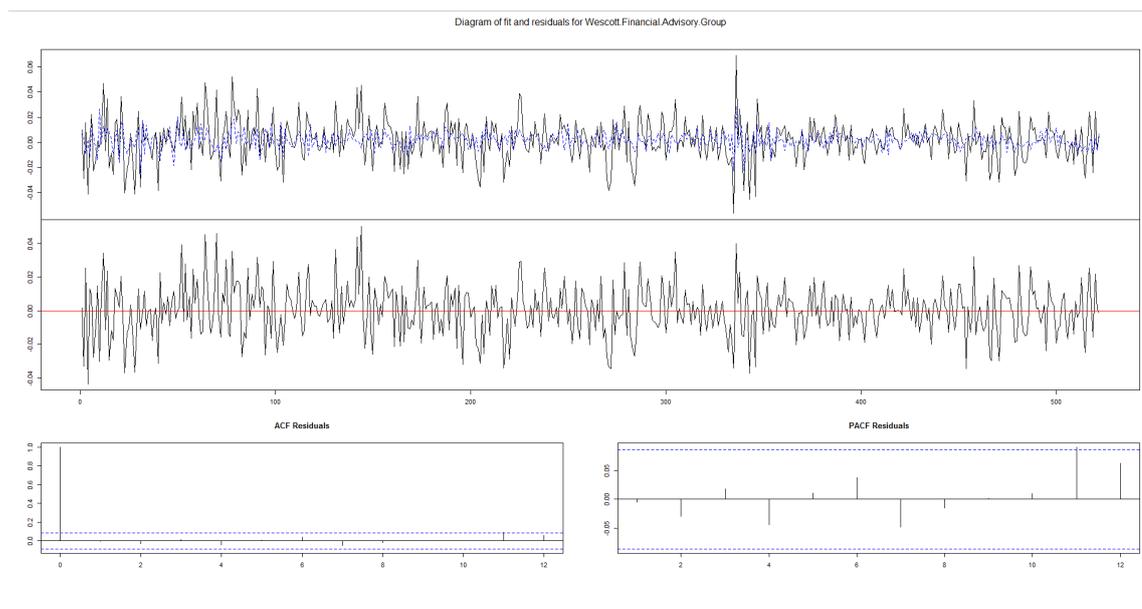

Figure 4.5: Diagram of Fit and Residuals of Sample Advisor

# Chapter 5

# Conclusions

This paper basically talks about analysis of investment advisors intelligence. A novel problem of discovering leaders from set of time series data based on lagged correlation has been proposed. A time series is learned as leader time series. The movement of leader time series triggers many other time series which are called as followers. Behavior of leader time series helps in learning the behavior of the followers time series. Proceeding with further analysis of investment advisors intelligence led to pattern recognition of the investment advisor behavior. An approach towards this problem showed great interest in using EAs for pattern recognition tasks, and also came with other possible use of EAs combined with other approaches for the development of fully automated pattern recognition systems